\documentclass[12pt]{iopart}

\usepackage{graphicx}
\usepackage[colorlinks=true]{hyperref}
\usepackage{amssymb}
\usepackage{longtable}
\usepackage{rotating}
\usepackage{color}
\usepackage{epsfig}
\usepackage{units}
\usepackage{wasysym}
\usepackage{subfigure}
\begin{document}

\title[Identification of long-duration noise transients in LIGO and Virgo]{Identification of long-duration noise transients in LIGO and Virgo}

\author{Michael~W.~Coughlin for the LIGO Scientific Collaboration and the Virgo Collaboration}

\address{Physics and Astronomy,
Carleton College,
Northfield, MN, 55057, USA}
\ead{coughlim@carleton.edu}

\begin{abstract}
The LIGO and Virgo detectors are sensitive to a variety of noise sources, such as instrumental artifacts and environmental disturbances. The Stochastic Transient Analysis Multi-detector Pipeline (STAMP) has been developed to search for long-duration (t$\gtrsim$1s) gravitational-wave (GW) signals. This pipeline can also be used to identify environmental noise transients. Here we present an algorithm to determine when long-duration noise sources couple into the interferometers, as well as identify what these noise sources are. We analyze the cross-power between a GW strain channel and an environmental sensor, using pattern recognition tools to identify statistically significant structure in cross-power time-frequency maps. We identify interferometer noise from airplanes, helicopters, thunderstorms and other sources. Examples from LIGO's sixth science run, S6, and Virgo's third scientific run, VSR3, are presented.
\end{abstract}

\maketitle

\section{Introduction}

The Laser Interferometer Gravitational-wave Observatory (LIGO) \cite{LIGO} and Virgo \cite{VIRGO} detector are part of a network of gravitational-wave (GW) interferometers seeking to make the first direct observations of GWs \footnote{As presented at the Gravitational-wave Physics and Astronomy Workshop in Milwaukee, Wisconsin, January 26-29, 2011.}. These detectors are designed to be sensitive to GWs emitted by the coalescence of neutron stars or black holes, core collapse supernovae, pulsars, or cosmological processes that happened during the early universe \cite{lrr-2009-2}.

In July 2009, LIGO commenced its sixth science run (S6), while Virgo started its second and third science runs (VSR2 and VSR3) interrupted between January 2010 and July 2010 for a detector upgrade. In this paper, only VSR3 data have been studied. LIGO consists of three Michelson interferometers at two different sites. There are two located in Hanford, Washington, known as H1 and H2, with 4km and 2km arms respectively, though only H1 acquired data during S6. The other is located in Livingston, LA, a 4km interferometer known as L1. Virgo consists of a 3km interferometer located in Cascina, Italy, known as V1. 


In order to make confident detections of GWs, it is necessary to understand detector noise artifacts. Despite their insulation, the detectors are susceptible to a variety of instrumental and environmental noise sources that decrease their detection sensitivity \cite{S6DetectorChar}. Short, non-astrophysical transient events or \emph{glitches} can mask or mimic real signals. Environmental noise can couple into the interferometer through mechanical vibration or because of electromagnetic influence. Seismic motion from human activity near the sites, from wind and from ocean waves are among the most common sources of these disturbances. Though not as common as ground traffic, airplanes and helicopters also affect the GW strain amplitude, or $h(t)$.
 
To mitigate the effects of noise artifacts in GW searches, periods of time when the detector noise level is unsuitable for GW detection are excluded or \emph{vetoed} \cite{S6DetectorChar}. It is necessary to develop techniques to veto non-astrophysical events both effectively and safely. A veto is considered \emph{safe} if no GW signal is dismissed. Vetoes reduce the adverse effect of noise on searches for GWs. Probes such as microphones and seismometers are placed at sensitive locations around the detectors. Hundreds of channels are thus recorded continuously. These are called Physical Environmental Monitoring (PEM)
in LIGO and auxiliary channels in Virgo. In the following, we will use PEM to refer to any of these channels, regardless of detector. A number of veto definition methods have been developed using these channels to pinpoint short duration noise transient sources \cite{CBCVeto,UPV,VirgoDQ}. Techniques for identifying narrow spectral features (``lines'') that could mimic a continuous GW source have been developed as well \cite{Fscan}.

The Stochastic Transient Analysis Multi-detector Pipeline (STAMP) was recently developed to search LIGO and Virgo data for long GW transients whose durations may range from $\cal{O}(\unit[]{s})$ to weeks. It is a MATLAB implementation of the excess cross-power statistic described here \cite{STAMP}. This algorithm uses frequency-time ($ft$)-maps of GW strain \emph{cross-power}, calculated by cross-correlating strain channels from two spatially separated detectors. Pattern recognition methods (including the box- and Radon algorithms) are then applied to identify GW signals. A complete discussion can be found in \cite{STAMP}.

\begin{figure*}[hbtp!]
 \centering
 \includegraphics[width=4in]{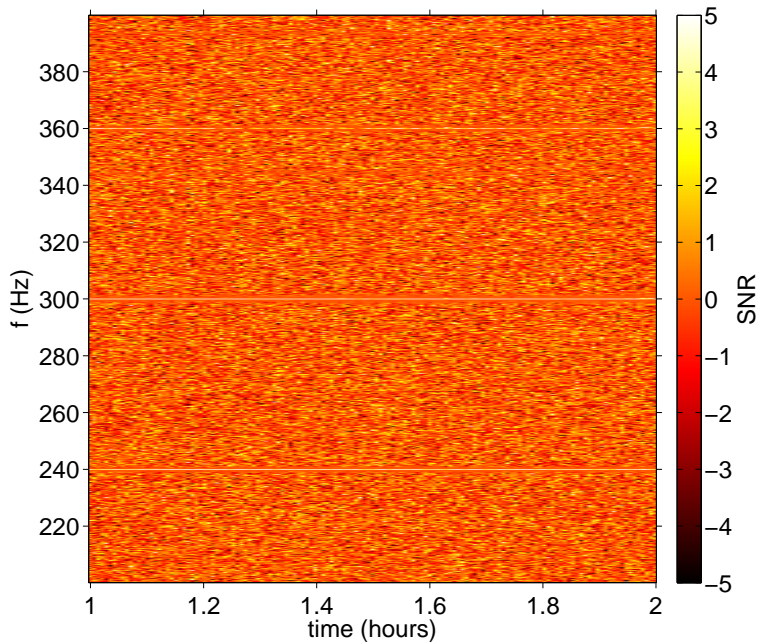}
 \caption{$ft$-map of the cross-power between the GW channel and an accelerometer at the Hanford site using 52s segments. There are strong lines at multiples of 60Hz, corresponding to the harmonics of the 60Hz power mains present at both sites.}
 \label{fig:StochMap}
\end{figure*}

Here we apply STAMP techniques to analyze cross-power $ft$-maps obtained with a PEM channel and $h(t)$ for H1, L1 and V1 data during S6 and VSR3 runs. By cross-correlating $h(t)$ with a PEM channel, there is no sensitivity to astrophysical signals --- which are not present in the PEM channels --- while preserving coherent noise artifacts. The same pattern recognition algorithms mentioned above are then used to find noise sources. To create the $ft$-maps, the cross-power is calculated in either 4s or 52s blocks over the total time segment requested, depending on the approximate duration of the type of noise source. We use overlapping Hann-windowed data. The advantage of this method is that only noise sources which couple into the GW strain amplitude are revealed. An excess of noise in a PEM channel alone will not induce excess cross-power when the PEM channel is cross-correlated with a GW strain channel. By looking for structure in these $ft$-maps (see Fig.~\ref{fig:StochMap} for an example), it is possible to identify the origin of the noise.

The remainder of the paper is organized as follows. In section 2, we present the search for airplane and helicopter noise artifacts. Identification of a noise source covering the Crab pulsar frequency during two weeks in S6 is discussed in section 3. In section 4, we present a search for thunderstorms. Our conclusions are discussed in section 5.

\section{Airplanes and Helicopters}

\subsection{Airplanes}

Airplanes are a source of non-stationary noise for LIGO and Virgo, as they generate acoustic noise. An airplane's movement with respect to the detector causes a Doppler-modulated signal on an $ft$-map made with acoustic PEM channels. The airplane's frequency (assuming a straight-line, constant-speed trajectory) falls monotonically with time. The existing LIGO airplane veto system (called planemon \cite{PlaneMon}) has been shown to flag airplanes observed in microphone channels, and these flags have been shown to agree with airplane flight data \cite{STAMP}. A disadvantage to this method is that it does not determine if the vibration from the airplanes couple into $h(t)$. To reduce misidentification, it also requires multiple microphones to be triggered to identify a signal, and thus planemon may miss low amplitude airplane noise that triggers only one microphone.

In order to use STAMP to identify airplane events, $ft$-maps of 4s segments correlating $h(t)$ with acoustic channels (microphones) are computed in 400s blocks. 400s was chosen so as to capture the entire airplane track, but not so long that the track loses its curvature in an $ft$-map. The average length for an airplane track during S6 and VSR3 was 115s. The frequency range of the $ft$-map was restricted to 65Hz-115Hz, as most of the airplane signal appears in this frequency band. We then take the absolute value of the $ft$-map, as transient noise artifacts can produce complex (not positive-definite) cross-power because the cross-power phase depends on the coupling of the environmental noise into $h(t)$. Purely horizontal and vertical lines in the maps, which can occur due to glitches and instrumental lines, are removed. 

\begin{figure*}[hbtp!]
\centering \subfigure{}
\includegraphics[width=3in]{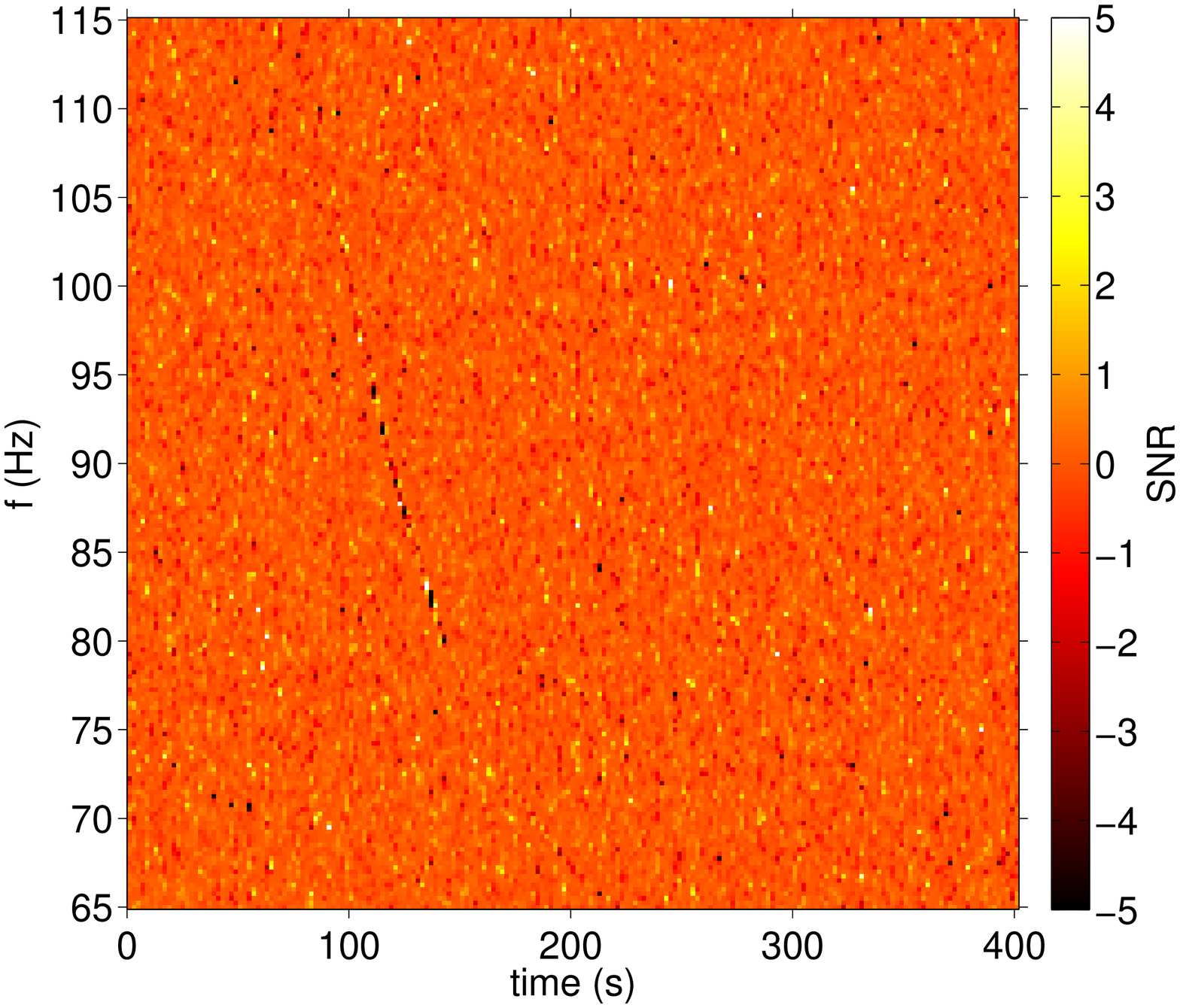}
\centering \subfigure{}
\includegraphics[width=3in]{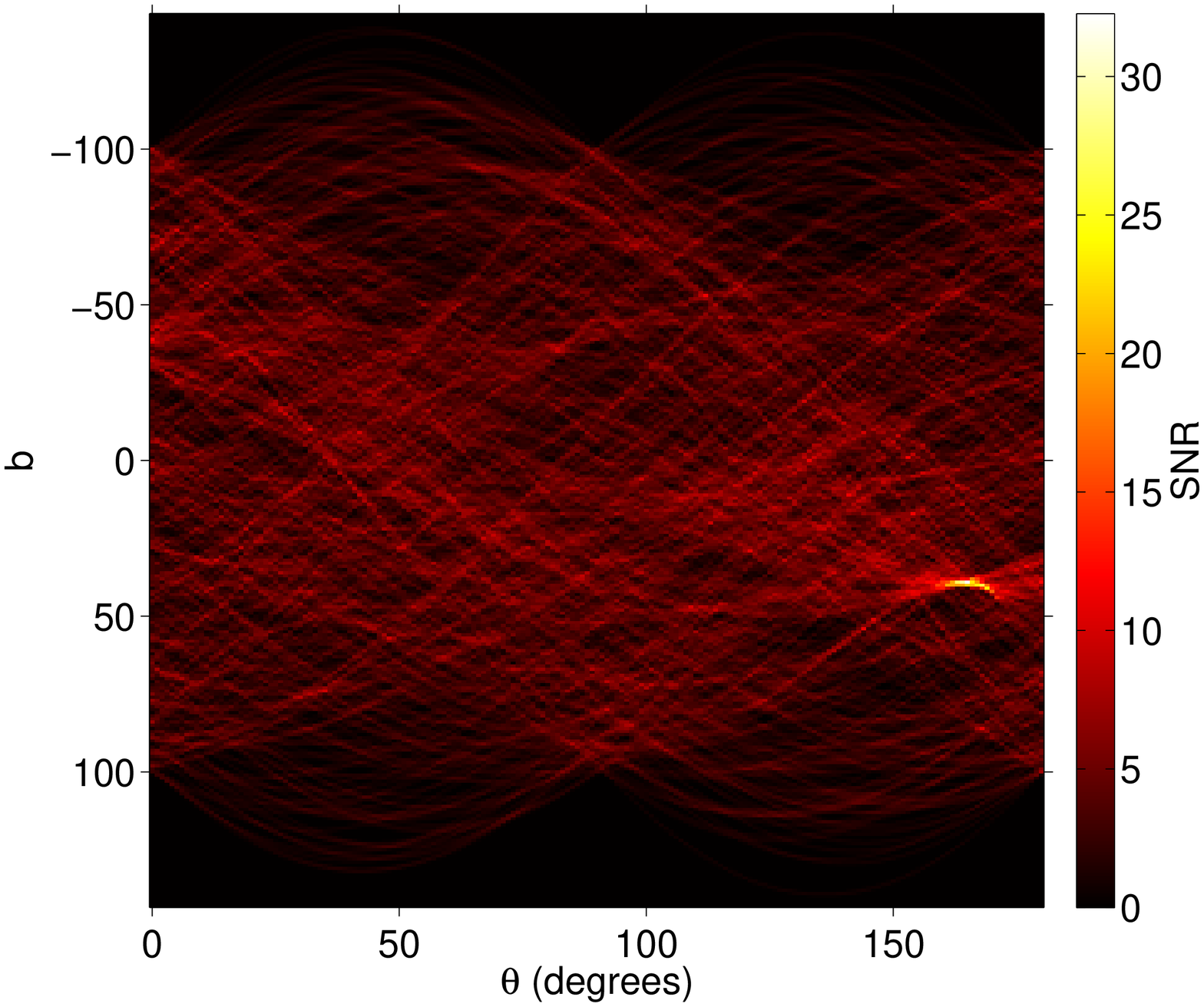}
\caption{Left: $ft$-map of the cross-power between the GW channel and an acoustic channel. The airplane track can be seen on the left side of the plot. Right: Radon map of this event, where the impact parameter b is the shortest distance to the line from the origin and $\theta$ is the angle from the horizontal. The bright spot on the right panel corresponds to the airplane's signal in Radon space.}
\label{fig:AirplaneStochRadon}
\end{figure*}

An example airplane signal can be seen in the $ft$-map on the left of Fig.~\ref{fig:AirplaneStochRadon}. We next apply a Radon transform \cite{Radon} to each $ft$-map, an example of which can be seen on the right of Fig.~\ref{fig:AirplaneStochRadon}. The Radon transform produces Radon SNR, which is a weighted sum of cross-power along lines parameterized by the impact parameter b, the shortest distance to the line from the origin and $\theta$, the angle from the horizontal. Although airplane tracks are not quite lines in $ft$-space, the approximation of the tracks as lines is suitable for a simple identification. We then record the maximum Radon SNR, denoted as $\mathrm{SNR}_{\Gamma}$, which is the value of the largest pixel in each Radon map.

\begin{figure*}[hbtp!]
\centering
\includegraphics[width=4in]{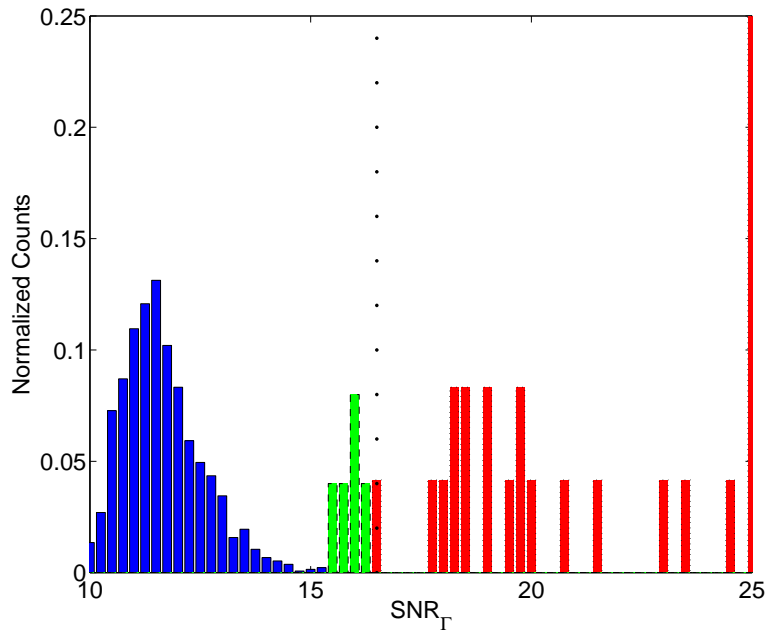}
\caption{Histogram of $\mathrm{SNR}_{\Gamma}$ for above threshold airplane events (in red, dotted outline), near threshold events (green dashed outline) and below-threshold background events (in blue, solid outline) in a week of
Virgo data. The dotted black line corresponds to the threshold of 16.5. The red, green and blue distributions are arbitrarily normalized for the purpose of plotting. About 93\% of the maps correspond to non-airplane noise, 6\% are airplanes, and 1\% are ambiguous. Entries with $\mathrm{SNR}_{\Gamma}=25$ record the maps with $\mathrm{SNR} \geq 25$.}
\label{fig:AirplaneHist}
\end{figure*}

To determine an approximate threshold for $\mathrm{SNR}_{\Gamma}$, a number of $ft$-maps for airplane tracks are checked by eye for each detector site. In Fig.~\ref{fig:AirplaneHist}, we show a histogram of $\mathrm{SNR}_{\Gamma}$ for the Virgo data. Using histograms like this, we set thresholds on $\mathrm{SNR}_{\Gamma}$. They are 16.5, 16, 16 for V1, H1 and L1 respectively.

\begin{figure*}[hbtp!]
\centering \subfigure{}
\includegraphics[width=3in]{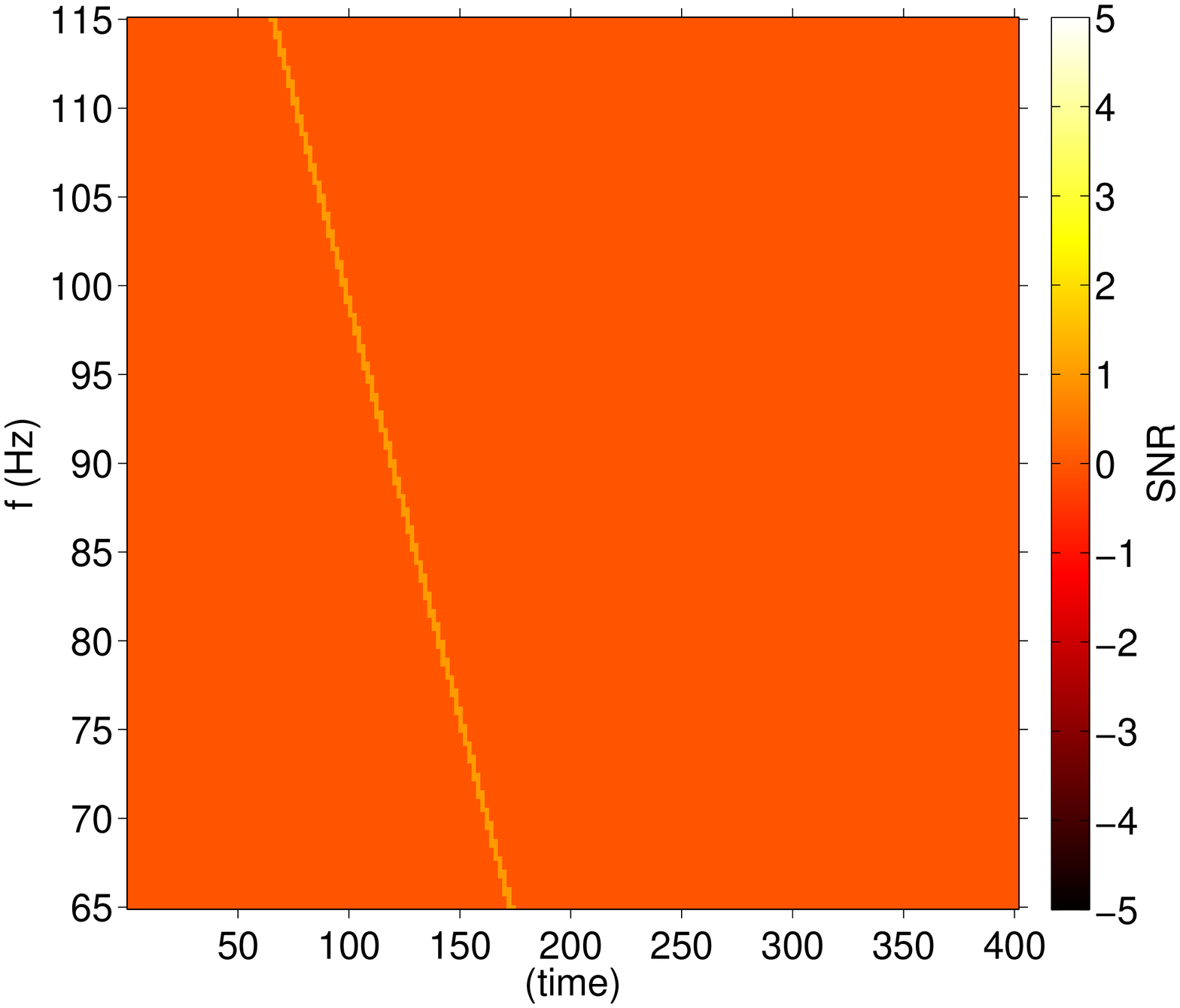}
\centering \subfigure{}
\includegraphics[width=3in]{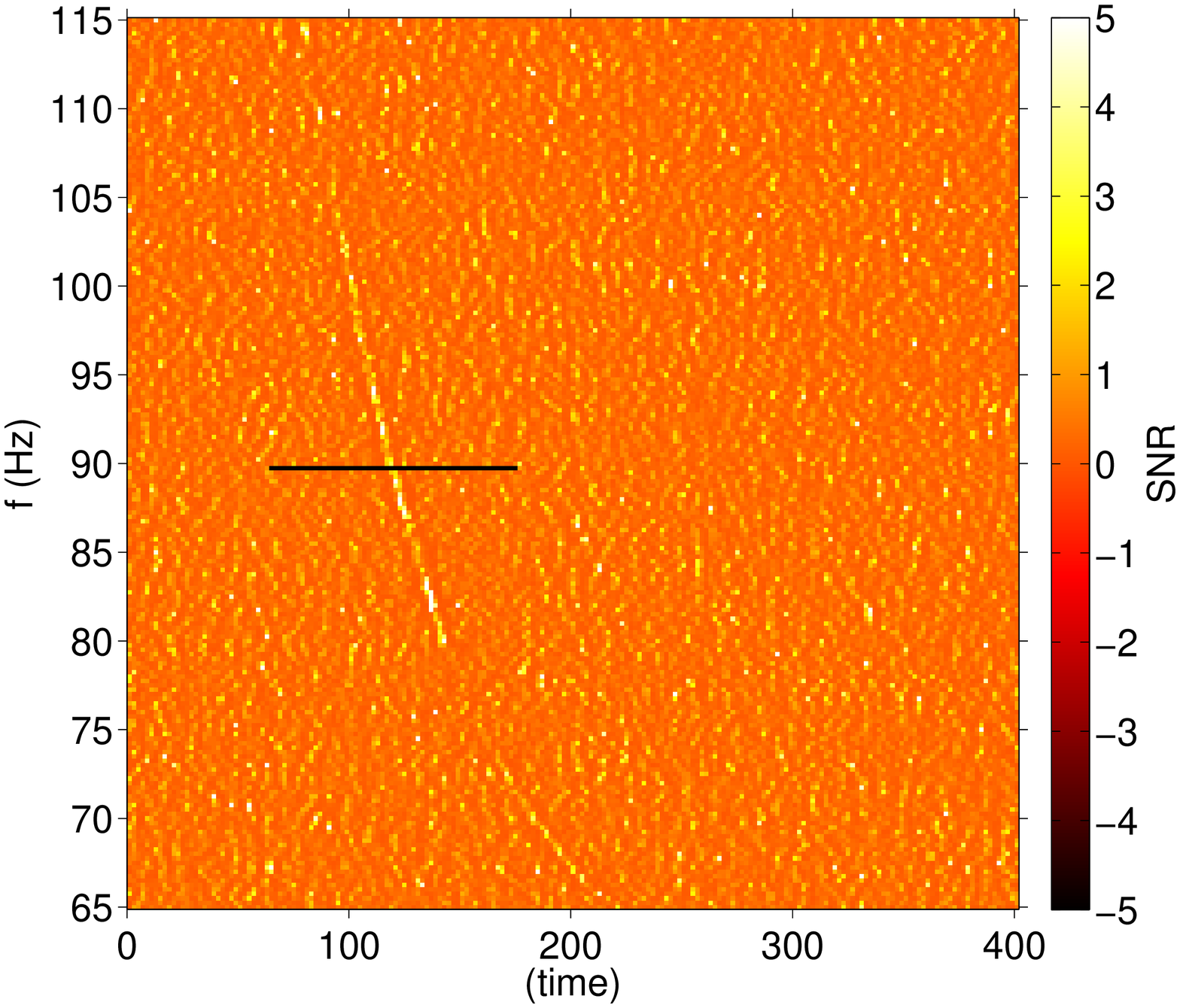}
\caption{Left: Radon reconstruction of the airplane event in Fig.~\ref{fig:AirplaneStochRadon}. Right: the ``veto'' window created with the reconstruction plot in Fig.~\ref{fig:AirplaneStochRadon}.}
\label{fig:AirplaneReconstructVeto}
\end{figure*}

For each map with $\mathrm{SNR}_{\Gamma}$ above the threshold, it is deemed to contain an airplane. To fit the track to a line, we set the Radon map to zero except where it is maximal and perform an inverse Radon transform back to $ft$-space (see Fig.~\ref{fig:AirplaneReconstructVeto} for an example). The start and stop times of the airplane noise in this frequency band are estimated to be the times at which the reconstructed track intersects the edge of the $ft$-map, and these define the veto window size for this airplane. 

This algorithm was run automatically for each day in VSR3 and S6d, which occurs in the last two months of S6. Each day was divided into 400s blocks, and $h(t)$ at each detector was correlated with the available microphones. For each block deemed to contain an airplane, the GPS times of the identified airplane were saved. For each day, a list of the times of the identified airplanes was compiled.

\begin{table}
  \begin{center}
  \scalebox{0.6}{
  \begin{tabular}{|c c c c c c|}
  \hline
  \small {Detector}& \small {Time Analyzed}& \small {Number of Analyzed Microphones}& \small {Number of Airplanes per Day}& \small {Dead Time}& \small {Vetoed HW Injections} \\
  \hline
  \small{H1}& \small{S6d}& \small{13}& \small{28}& \small{0.08\%}& \small{4/9752} \\
  \small{L1}& \small{S6d}& \small{9}& \small{21}& \small{0.14\%}& \small{9/8464} \\
  \small{V1}& \small{VSR3}& \small{4}& \small{2.9}& \small{0.09\%}& \small{0/3083} \\
  \hline
  \end{tabular}
  }
  \end{center}
 \caption{Results from the airplane analysis for VSR3 and S6d. For each detector, the time period analyzed, number of analyzed microphones, number of airplanes per day, dead time, and vetoed HW injections are listed. Numbers of airplanes per day detected have been cross-checked with other monitors.}
 \label{fig:Airplanes}
\end{table}

Over the entire run, to estimate the effect of airplanes in the detectors, the deadtime (percentage of data excluded by the airplane flag) was calculated. In order to determine if the veto is ``safe,'' we use simulated GW waveforms of burst and coalescing binary inspiral signals injected into the interferometer by applying a force on one of the mirrors of the Fabry-Perot cavity in one of the interferometer's arms to make sure that the veto is not issued by a GW event \cite{CBCVeto}. These results are listed in Table~\ref{fig:Airplanes}. For all three detectors, the number of vetoed injections is consistent with statistical expectations for a safe veto. Even so, a number of airplanes were found to be accidentally coincident with hardware injections. Due to the way vetoes are currently used, these periods would have been vetoed. But STAMP is able to determine the frequency band of the noise disturbance which can be well separated from a real GW signal frequency. It is possible to observe a GW event in the presence of an airplane (as long as the GW and airplane events do not overlap in the $ft$-map). It therefore seems plausible that during future science runs, this capability will be implemented to veto less data from GW analysis.

\subsection{Helicopters}

\begin{figure*}[hbtp!]
\centering \subfigure{}
\includegraphics[width=3in]{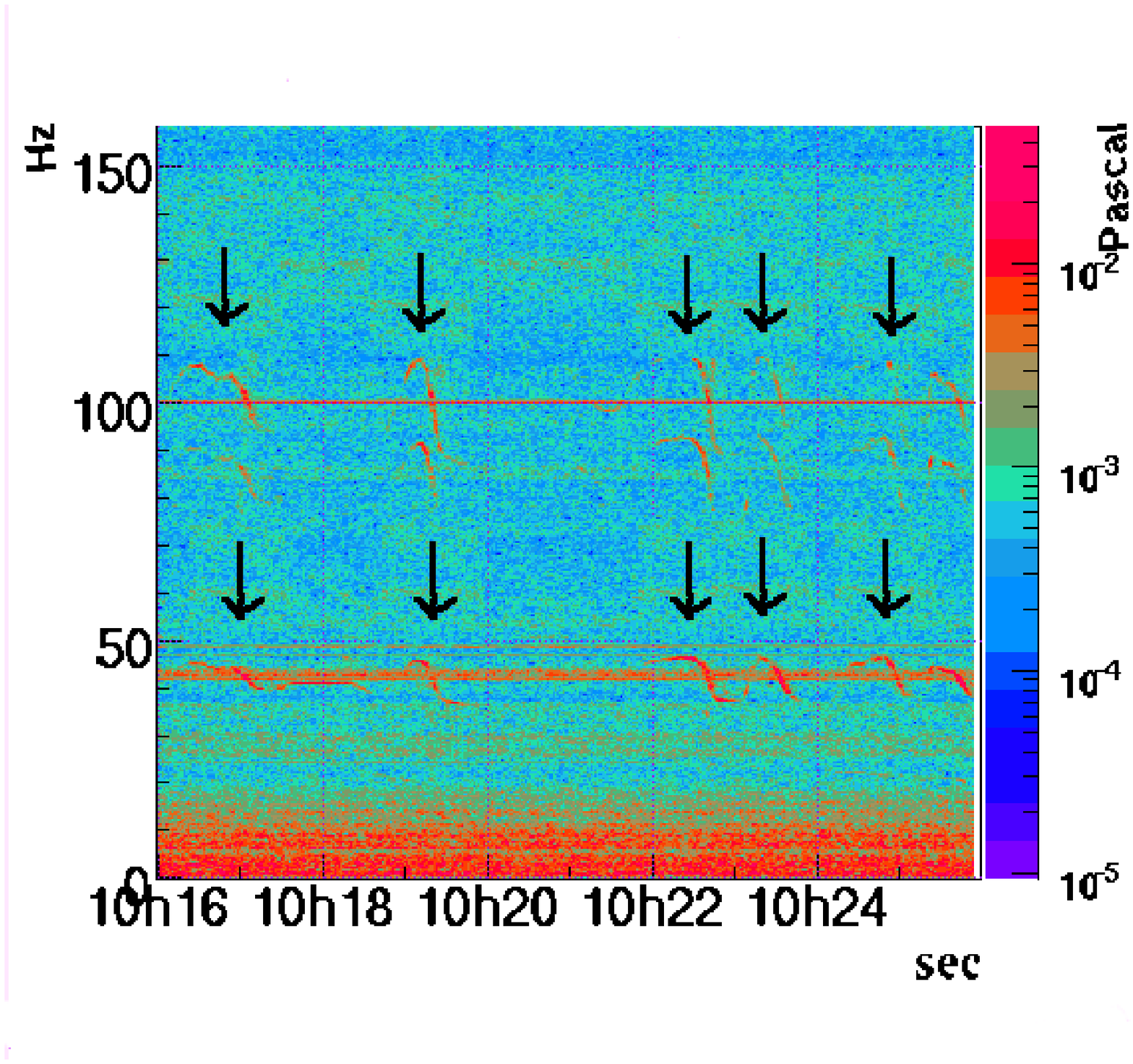}
\centering \subfigure{}
\includegraphics[width=3in]{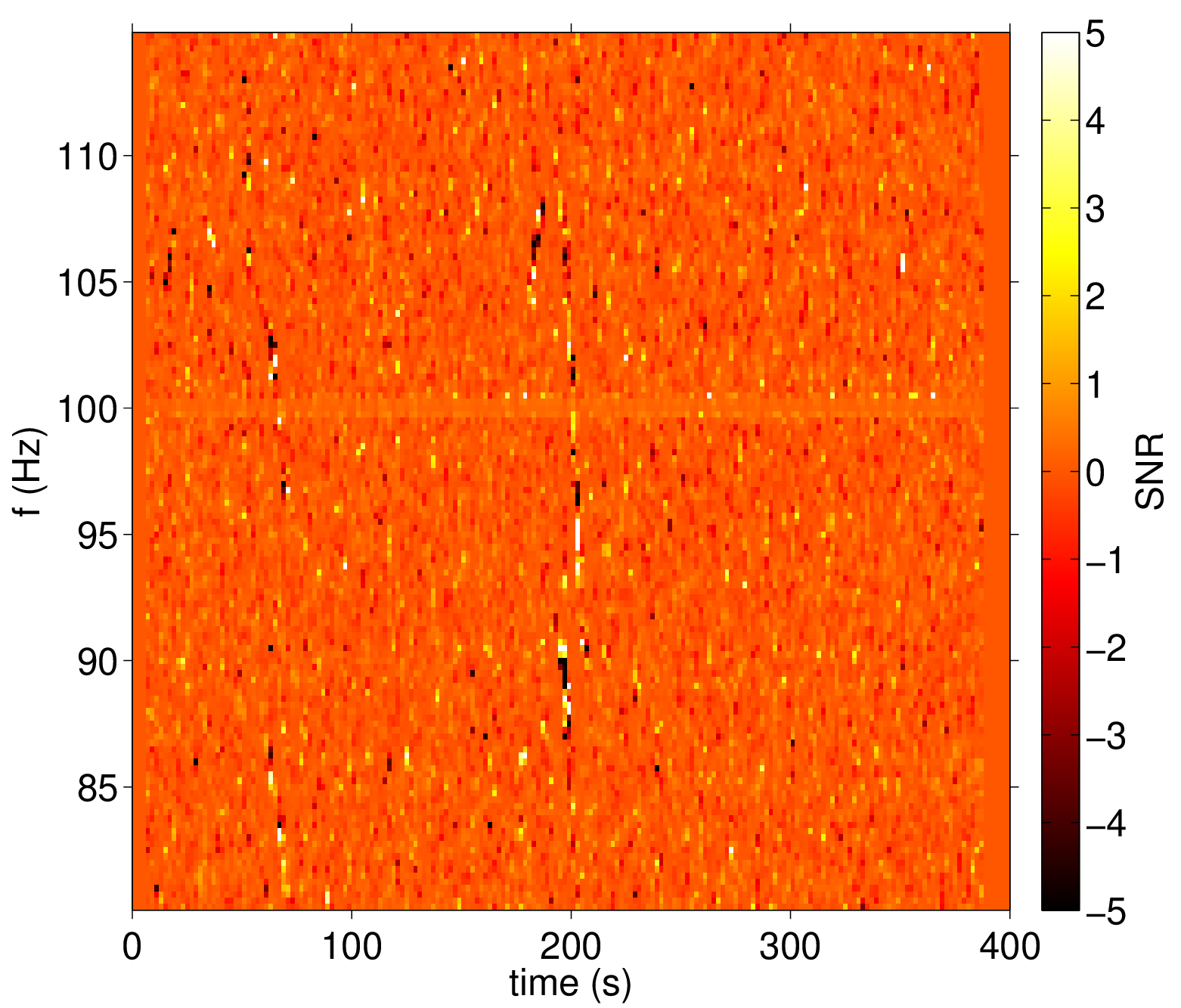}
\caption{Left: Spectrogram of a microphone in the central building of the Virgo detector at a time when a helicopter was flying near to the interferometer. The arrows indicate instances of helicopter noise. Right: $ft$-map of the cross-power between the GW channel and this microphone at this time. This plot shows that the noise of the helicopter is coupling with $h(t)$.}
\label{fig:Helicopters}
\end{figure*}

During VSR3, helicopters were a noise source for the Virgo detector, with tracks similar to those of airplanes, although their tracks contained more visible harmonics (see Fig.~\ref{fig:Helicopters} for an example). Some of the flybys were close enough to Virgo to degrade detector sensitivity dramatically. Due to their similar noise structure, the airplane algorithm identified a number of helicopter events which were vetoed.

\section{Identification of line near the Crab Frequency}

\begin{figure*}[hbtp!]
 \centering \subfigure{}
 \includegraphics[height=3in,width=3in,clip]{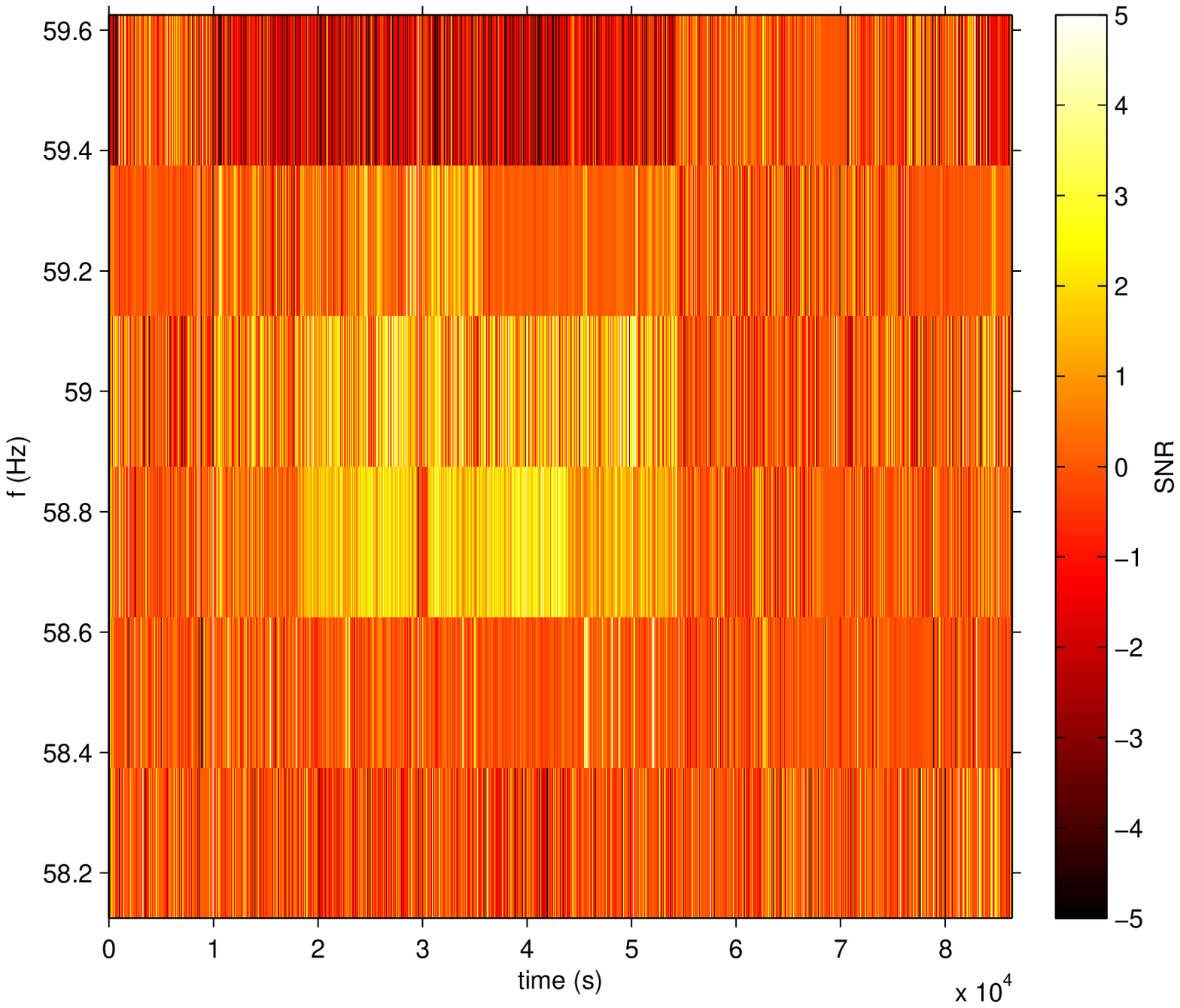}
 \centering \subfigure{}
 \includegraphics[height=3in,width=3in,clip]{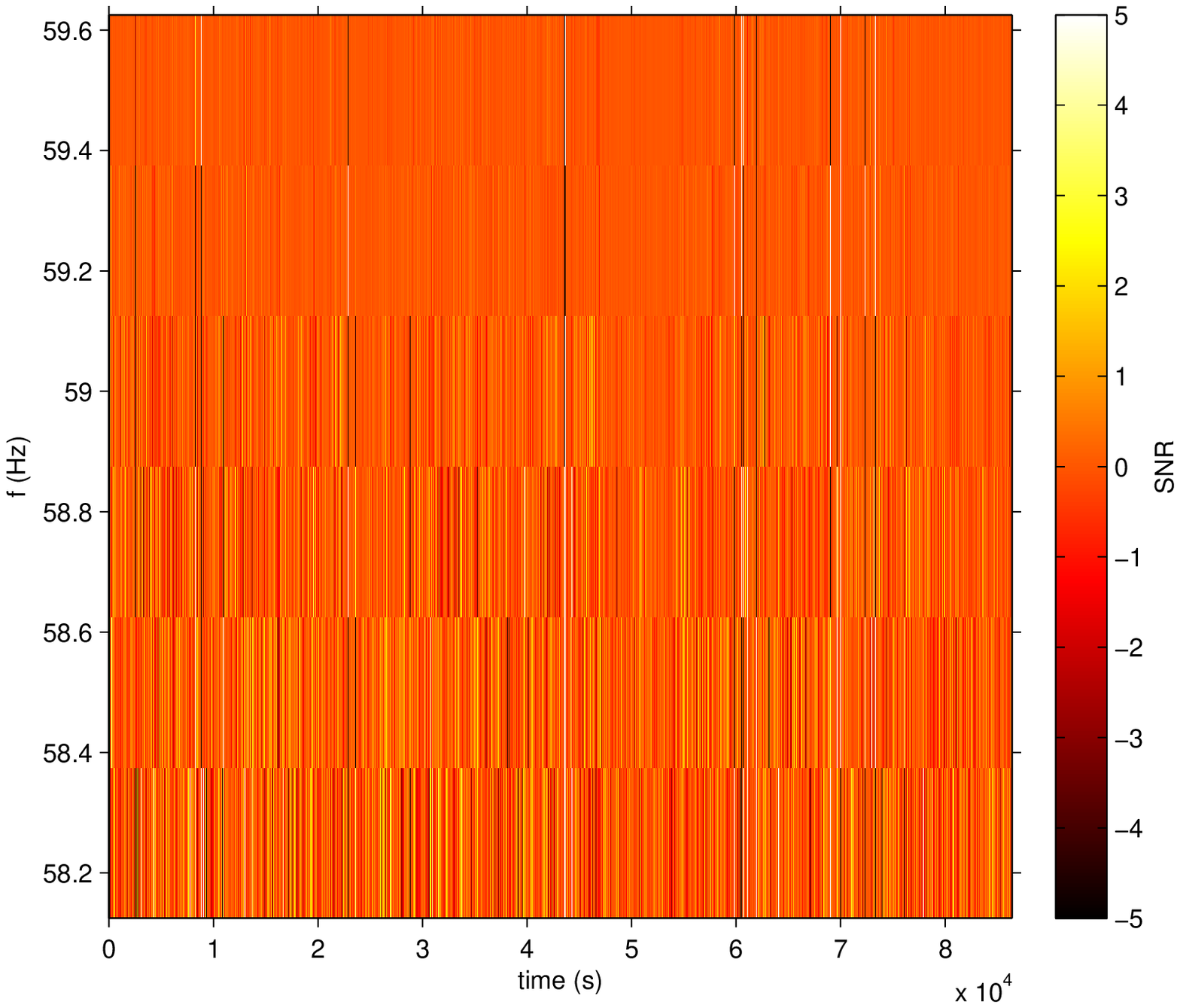}
 \caption{Left: $ft$-map of the cross-power between the GW channel and an accelerometer near to the water pump on June 8, 2010 (before the normal load was resumed). Please note that the line was not stationary and so the noise does not appear continuously in the $ft$-map. Right: $ft$-map of the same on June 13, 2010 (after the normal load was resumed).}
 \label{fig:CrabNoise}
\end{figure*}

The Crab pulsar, which has a GW signal frequency of approximately 59.55Hz, is an important target for LIGO and Virgo's continuous GW searches \cite{Crab}. During two weeks of LIGO's S6 run, an initially unidentified noise source appeared in $h(t)$ near the Crab frequency. $ft$-maps correlating $h(t)$ and all available PEM channels, using 52s segments, were calculated for this time. The noise was observed in a number of seismometers located inside of the central building, where the beam splitter and two of the test masses reside. A chilled water booster pump in the mechanical room was found to be vibrating at this frequency, coupling into $h(t)$. A number of weeks before, the pump's load had been reduced, causing its frequency to approach the Crab frequency. Subsequently, the load was returned to normal, eliminating the noise, as can be seen in Fig.~\ref{fig:CrabNoise}.

\section{Thunderstorms}
   
\begin{figure*}[hbtp!]
\centering \subfigure{}
\includegraphics[width=3in]{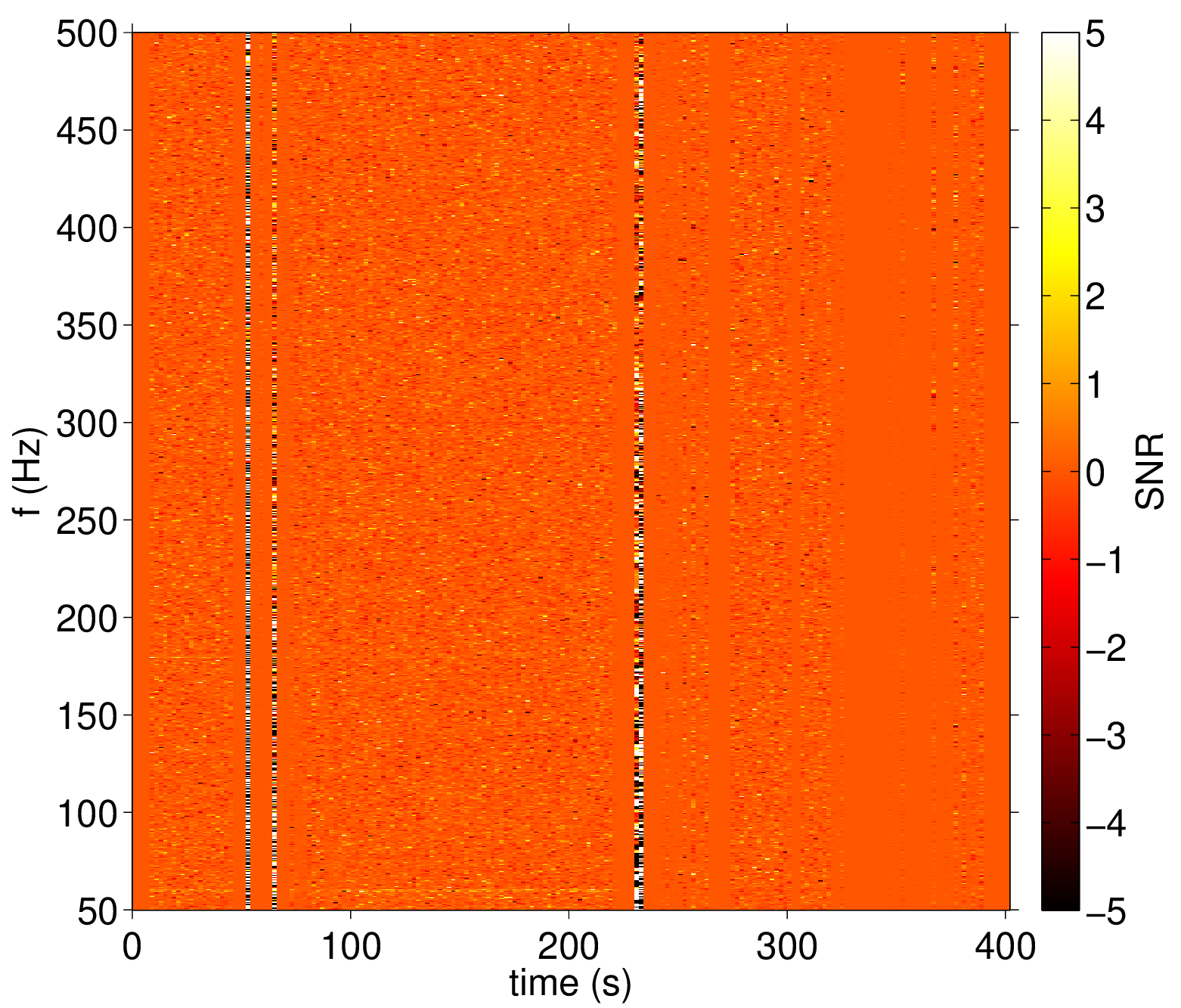}
\centering \subfigure{}
\includegraphics[width=3in]{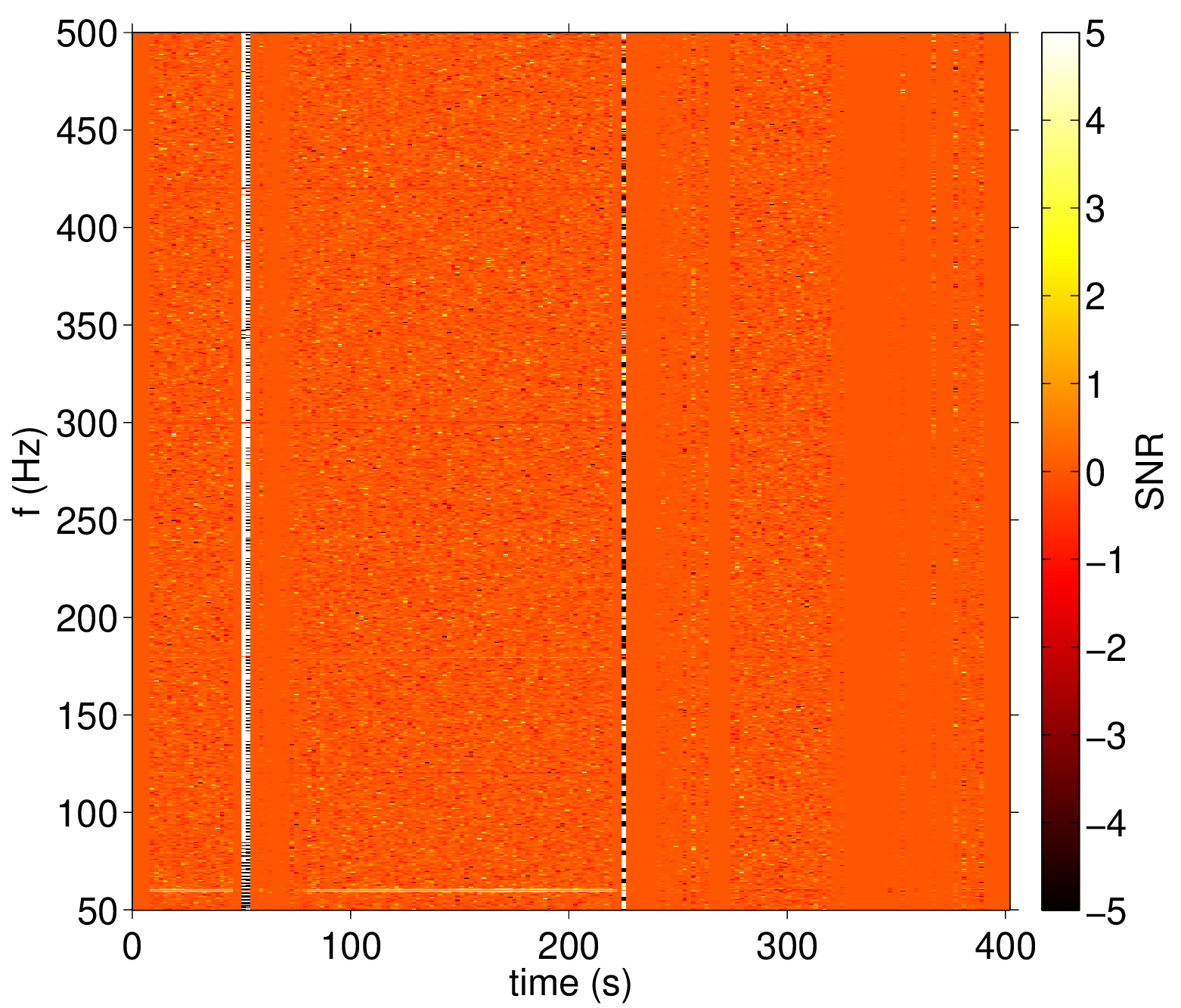}
\caption{Left: $ft$-map of the cross-power between the GW channel and a microphone during a coincident event at H1. Right: $ft$-map of the cross-power between the GW channel and a magnetometer at that time.}
\label{fig:Lightning}
\end{figure*}

Thunderstorms are a source of magnetic and acoustic environmental noise. They can prevent the detectors from running stably. In order to identify thunderstorms, we create $ft$-maps correlating the GW channel and microphones, as well as maps of the GW channel and magnetometers. We then search for broadband events of high cross-power in these maps. The times of these events are recorded for both the microphones and magnetometers, and events coincident in both the microphones and magnetometers within a 10s window are flagged as thunderstorm candidates (see Fig.~\ref{fig:Lightning} for an example). During S6d, 7 events were identified at Hanford, while 26 were identified at Livingston. During VSR3, 1 event was identified at Virgo.

To test the efficiency of the algorithm, the identified events were compared with 206 identified lightning events one night at H1 using a camera on the roof. During this time, only 2 of these lightning flashes were identified by the algorithm, resulting in a low efficiency of 1\%. To ensure that these were not accidental coincidences, we gave the coincident events artificial time offsets, or ``time slides,'' looking for lightning events. Time slides between -100s-100s in 20s intervals were performed, resulting in zero accidental coincidences. This suggests that the algorithm is detecting actual lightning flashes, but at a very low rate. There are a few possible reasons for this. One is that it is unlikely that each lightning strike has enough strength to appear in a microphone, magnetometer, and the interferometer strain amplitude. Also, it is likely that instead of detecting the magnetic fields of the lightning, we are detecting magnetic fields from power system glitches due to lighting strikes that hit power lines or other power distribution facilities. This was confirmed by observing events in voltage monitors on site coincident with the lightning events.

\section{Conclusions}
 
We have realized many noise hunting applications with the Stochastic Transient Analysis Multi-detector Pipeline (STAMP), and others are still in development. STAMP data quality flags were generated for use by LIGO and Virgo inspiral \cite{LIGOCBC,VirgoCBC} and burst \cite{LIGOBurst,VirgoBurst} GW searches. Looking ahead towards advanced detectors, STAMP will be an important tool for identifying not only the noise sources currently affecting the detector but also the new noise sources that will inevitably appear.

\ack

This article has been assigned LIGO Document Number P1100064. This project is funded by NSF Grant PHY-0854790 and the Kolenkow-Reitz Fund at Carleton. The authors gratefully acknowledge the support of the United States National Science Foundation for the construction and operation of the LIGO Laboratory, the Science and Technology Facilities Council of the United Kingdom, the Max-Planck-Society, and the State of Niedersachsen/Germany for support of the construction and operation of the GEO600 detector, and the Italian Istituto Nazionale di Fisica Nucleare and the French Centre National de la Recherche Scientifique for the construction and operation of the Virgo detector. The authors also gratefully acknowledge the support of the research by these agencies and by the Australian Research Council, the Council of Scientific and Industrial Research of India, the Istituto Nazionale di Fisica Nucleare of Italy, the Spanish Ministerio de Educaci\'on y Ciencia, the Conselleria d'Economia Hisenda i Innovaci\'o of the Govern de les Illes Balears, the Foundation for Fundamental Research on Matter supported by the Netherlands Organisation for Scientific Research, the Polish Ministry of Science and Higher Education, the FOCUS Programme of Foundation for Polish Science, the Royal Society, the Scottish Funding Council, the Scottish Universities Physics Alliance, The National Aeronautics and Space Administration, the Carnegie Trust, the Leverhulme Trust, the David and Lucile Packard Foundation, the Research Corporation, and the Alfred P. Sloan Foundation.

\section*{References}
\renewcommand\refname{Bibliography}
\bibliographystyle{iopart-num}
\bibliography{STAMPRef}

\providecommand{\newblock}{}
\begin{thebibliography}{10}
\expandafter\ifx\csname url\endcsname\relax
  \def\url#1{{\tt #1}}\fi
\expandafter\ifx\csname urlprefix\endcsname\relax\def\urlprefix{URL }\fi
\providecommand{\eprint}[2][]{\url{#2}}

\bibitem{LIGO}
{Abbott B et al} (LIGO Scientific Collaboration) 2009 {\em Reports on Progress
  in Physics\/} {\bf 72} 076901

\bibitem{VIRGO}
{Acernese F et al} (Virgo Collaboration) 2008 {\em Classical and Quantum
  Gravity\/} {\bf 25} 114045

\bibitem{lrr-2009-2}
{Sathyaprakash B S and Schutz B} 2009 {\em Living Reviews in Relativity\/} {\bf
  12} \urlprefix\url{http://www.livingreviews.org/lrr-2009-2}

\bibitem{S6DetectorChar}
{Christensen N for the LIGO Scientific Collaboration and the Virgo
  Collaboration} 2010 {\em Class. Quantum Grav.\/} {\bf 27} 194010

\bibitem{CBCVeto}
{Slutsky J et al} 2010 {\em Class. Quantum Grav.\/} {\bf 27} 165022

\bibitem{UPV}
{Isogai T for the LIGO Scientific Collaboration and the Virgo Collaboration}
  2010 {\em J. Phys.: Conf. Ser.\/} {\bf 243} 012005

\bibitem{VirgoDQ}
{Robinet F for the LIGO Scientific Collaboration and the Virgo Collaboration}
  2010 {\em Classical and Quantum Gravity\/} {\bf 27} 194012

\bibitem{Fscan}
{Coughlin M for the LIGO Scientific Collaboration and the Virgo Collaboration}
  2010 {\em J. Phys.: Conf. Ser.\/} {\bf 243} 012010

\bibitem{STAMP}
{Thrane E et al} 2011 {\em Phys. Rev. D\/} {\bf 83} 083004

\bibitem{PlaneMon}
Goetz E and Riles K 2010 {\em LIGO DCC T050174-00-D\/}

\bibitem{Radon}
Deans S~R 2007 {\em The Radon Transform and Some of Its Applications\/} 2nd ed
  (Dover Publications, Inc.)

\bibitem{Crab}
{Abbott B et al} (LIGO Scientific Collaboration) 2008 {\em ApJ\/} {\bf 683} 1

\bibitem{LIGOCBC}
{Abbott B et al} (LIGO Scientific Collaboration) 2009 {\em Phys. Rev. D\/} {\bf
  80} 047101

\bibitem{VirgoCBC}
{Acernese F et al} (Virgo Collaboration) 2007 {\em Class. Quantum Grav.\/} {\bf
  24} 5767

\bibitem{LIGOBurst}
{Abbott B et al} (LIGO Scientific Collaboration) 2009 {\em Phys. Rev. D\/} {\bf
  80} 102001

\bibitem{VirgoBurst}
{Acernese F et al} (Virgo Collaboration) 2009 {\em Class. Quantum Grav.\/} {\bf
  26} 085009

\end{thebibliography}

\end{document}